\documentclass[12pt]{article}
\pdfoutput=1
\usepackage{amsmath,amsfonts,graphicx,color,bbm,tikz,bm,setspace}
\usepackage{comment}
\usepackage[nosort]{cite}
\usepackage{subfigure}
\usepackage{qcircuit}
\usepackage{multirow}
\usetikzlibrary{calc,positioning}
\usetikzlibrary{patterns,arrows,decorations.pathreplacing}
\usepackage{caption}
\usepackage{ulem}
\tikzset{>=stealth}
\usepackage{lipsum}
\usepackage{tabularx}
\usepackage{fullpage}
\usepackage{hyperref}
\usepackage{bbold}
\usepackage[english]{babel}
\newtheorem{definition}{Definition}[section]

\newtheorem{remark}{Remark}[section]

\usepackage{listings}
\usepackage{tcolorbox}

\newcommand{\mathsym}[1]{{}}
\newcommand{\unicode}[1]{{}}

\makeatletter\@addtoreset{equation}{section}\makeatother

\newcommand{\be}{\begin{equation}}
\newcommand{\ee}{\end{equation}}
\def\beq{\begin{equation}}
\def\eeq{\end{equation}}
\newcommand{\bea}{\begin{eqnarray}}
\newcommand{\eea}{\end{eqnarray}}

\def\nn{\nonumber}

\renewcommand{\title}[1]{\vbox{\center\LARGE{#1}}\vspace{3mm}}
\renewcommand{\author}[1]{\vbox{\center{#1}}\vspace{3mm}}

\newcommand{\email}[1]{\vbox{\center\tt#1}\vspace{3mm}}

\begin{document}
%\begin{titlepage}

\begin{center}
%{\large {\bf New constant 4 by 4 Yang-Baxter solutions}}
%{\large {\bf Algebraic methods to reproduce Hietarinta's classification of 4 by 4 constant Yang-Baxter operators }}
{\large {\bf Algebraic classification of Hietarinta's solutions of Yang-Baxter equations~:~invertible $4\times 4$ operators }}

\author{Somnath Maity,$^a$ Vivek Kumar Singh,$^b$ Pramod Padmanabhan,$^a$ Vladimir Korepin$^{c}$}

%\vskip 0.25cm
{$^a${\it School of Basic Sciences,\\ Indian Institute of Technology, Bhubaneswar, 752050, India}}
\vskip0.1cm
{$^b${\it Center for Quantum and Topological Systems (CQTS), NYUAD Research Institute,\\ New York University Abu Dhabi, PO Box 129188, Abu Dhabi, UAE}}
\vskip0.1cm
{$^c${\it C. N. Yang Institute for Theoretical Physics, \\ Stony Brook University, New York 11794, USA}}

\email{somnathmaity126@gmail.com, vks2024@nyu.edu, pramod23phys@gmail.com, vladimir.korepin@stonybrook.edu}

\vskip 0.5cm 

\end{center}

%%%%%%%%%%%

\abstract{
\noindent 
In order to examine the simulation of integrable quantum systems using quantum computers, it is crucial to first classify Yang-Baxter operators. Hietarinta was among the first to classify constant Yang-Baxter solutions for a two-dimensional local Hilbert space (qubit representation). Including the one produced by the permutation operator, he was able to construct eleven families of invertible solutions. These techniques are effective for 4 by 4 solutions, but they become difficult to use for representations with more dimensions. To get over this limitation, we use algebraic ans\"{a}tze to generate the constant Yang-Baxter solutions in a representation independent way. We employ four distinct algebraic structures that, depending on the qubit representation, replicate 10 of the 11 Hietarinta families. Among the techniques are partition algebras, Clifford algebras, Temperley-Lieb algebras, and a collection of commuting operators. Using these techniques, we do not obtain the $(2,2)$ Hietarinta class.

}

%\end{titlepage}
%\tableofcontents 

%%%%%%%%%%%%%%%%%%%%%%%%%%
\section{Introduction}
\label{sec:Introduction}
%%%%%%%%%%%%%%%%%%%%%%%%%%
Through the algebraic Bethe ansatz \cite{slavnov2019algebraicbetheansatz}, the Yang-Baxter equation (YBE) serves as the foundation for two-dimensional integrable models \cite{YangCN1967,BAXTER1972193,Baxter1982ExactlySM} and the quantum inverse scattering method \cite{takhtadzhyan1979,Takhtadzhan_1979,Korepin1993QuantumIS}. This construction is based on its solutions, the Yang-Baxter operators (YBO's). They have several applications. For instance, the creation of knot and link polynomials relies heavily on their spectral parameter independent versions, called constant YBOs, which are braid group generators \cite{Akutsu:1987bc,Akutsu:1987ea,Akutsu:1987qs,kauffman-knots,Turaev1988TheYE}. YBOs have been used more recently as quantum gates that enable universal quantum computation \cite{kauffman2004braiding,zhang2005universal,zhang2005yang,zhang2008braidgrouptemperleyliebalgebra,zhang2024geometricrepresentationsbraidyangbaxter}. In order to benchmark the latter, these operators are utilized to simulate integrable quantum models on quantum computers \cite{aleiner2021bethe,Sopena_2022}. Therefore, it is crucial to identify systematic ways for classifying these operators because of their wide range of applications in mathematics and physics.

In \cite{Sogo,Hlavaty_1987,Fei:1991zn,Zhang_1991}, some of the first attempts to classify both spectral parameter-dependent and constant YBOs are presented. In a series of papers, Hietarinta offered a comprehensive classification of invertible 4 by 4 constant YBOs \cite{HIETARINTA-PLA,hietarinta1993-JMP-Long,hietarinta-conferenceProceeding,Hietarinta1993-BookChapter}. These projects were restricted to building 4 by 4 YBOs. In their search for answers, they simplify the matrix analysis by taking advantage of the YBE's symmetries. For small-sized matrices, this is helpful in solving highly non-linear systems of equations. As a result, the final answers depend on the representation. 

Finding techniques that solve the YBE in a representation-independent way that yield solutions of arbitrary matrix sizes is therefore desirable. By building algebraic solutions of the YBE utilizing four different kinds of algebraic structures, we take a step in that direction. Clifford algebras, Temperley-Lieb algebras \cite{Temperley:1971iq,temperley2004relations}, a set of commuting operators, and partition algebras \cite{Martin1993algebraic,Martin1994temperley,Martin1996structure,halverson2004partitionalgebras} are among them. When the two-dimensional representation is chosen, we recover all 11 Hietarinta classes, with the exception of the $(2,2)$ class. 

The remainder of this work is structured as follows. The classification of Hietarinta is reviewed in Sec. \ref{sec:reviewH}. In order to define an equivalency class of YBOs, this section also examines the symmetries of the YBE. The various algebraic structures that will be employed to build the algebraic solutions in Sec. \ref{sec:algebraicStructures} are then described. Sec. \ref{sec:algebraicHietarinta} contains the key results of this article, including all the new algebraic solutions of the YBE. The results are summarized in Table \ref{tab:summary}. In Sec. \ref{sec:conclusion}, we wrap up with a brief conclusion and suggestions for further research.

%%%%%%%%%%%%%%%%%%%%%%%%%%%%%%%%%%%%%%%%%%%%%%%%%
\section{Hietarinta's classification}
\label{sec:reviewH}
%%%%%%%%%%%%%%%%%%%%%%%%%%%%%%%%%%%%%%%%%%%%%%%%%
We review the classification of 4 by 4 constant YBO's by Hietarinta \cite{HIETARINTA-PLA,hietarinta1993-JMP-Long,hietarinta-conferenceProceeding,Hietarinta1993-BookChapter}. We consider just the invertible solutions. The notation for the rest of the paper is also fixed in the process. 

The YBE can be written in two equivalent forms. The first form is adapted from relations satisfied by generators of the braid group. This is the braided form of the YBE:
\begin{equation}\label{eq:YBEbraided}
    \tilde{R}_{12}\tilde{R}_{23}\tilde{R}_{12} = \tilde{R}_{23}\tilde{R}_{12}\tilde{R}_{23}.
\end{equation}
The knot theory literature frequently uses this form \cite{kauffman-knots,Akutsu:1987bc,Akutsu:1987qs,Akutsu:1987ea}. The local Hilbert spaces ($V$) that the operators $R$ act on are indicated by the indices that appear in this equation. It should be noted that these operators can generally be implemented on Hilbert spaces of finite and infinite dimensions. In order to concentrate on matrix representations of the operator $R$, we shall assume that $V$ is finite-dimensional in this study. On $V\otimes V$, the $R$-matrices act non-trivially.  

The form of the YBE that finds use in the integrable model literature is the non-braided form or the vertex form:
\begin{equation}\label{eq:YBEnonbraided}
    R_{12}R_{13}R_{23} = R_{23}R_{13}R_{12}.
\end{equation}
It is important to observe the distinction in the index structure of the two versions of the YBE. Each of the three indices occurs twice on both sides of the equation in the non-braided version \eqref{eq:YBEnonbraided}. This isn't the case with the YBE's braided form \eqref{eq:YBEbraided}. In Sec. \ref{sec:algebraicHietarinta}, we will use this property to generate some very straightforward solutions to the YBE in \eqref{eq:YBEnonbraided}.  The YBEs are equations on $V\otimes V\otimes V$ in both forms.

The solutions to the two forms of the YBE are in one-to-one correspondence as:
\begin{equation}
    \tilde{R} = PR,
\end{equation}
where $P$ is the permutation operator on $V\otimes V$. The $M-1$ permutation operators $P_i\equiv P_{i, i+1}$, for $i\in\{1,2,\cdots,M-1\}$, generate the permutation group $S_M$. They satisfy the relations
\begin{eqnarray}\label{eq:permRelations}
    P_{i}P_{i+1}P_i = P_{i+1}P_iP_{i+1};~P_i^2=\mathbb{1};~P_iP_j=P_jP_i~\textrm{when}~|i-j|>1.
\end{eqnarray}
When $V=\mathbb{C}^2$, it can be written in terms of the Pauli matrices ($X$, $Y$ and $Z$) as:
\begin{equation}\label{eq:permutationC2}
   P = \frac{1}{2}\left[\mathbb{1} + X\otimes X + Y\otimes Y + Z\otimes Z \right] = \begin{pmatrix}
       1 & 0 & 0 & 0 \\
       0 & 0 & 1 & 0 \\
       0 & 1 & 0 & 0 \\
       0 & 0 & 0 & 1
   \end{pmatrix}. 
\end{equation}

%%%%%%%%%%%%%%%%%%%%%%%%%%%%%%%%%%%%%%%%%%%%%%%%%
\subsection{Symmetries of the YBE}
\label{subsec:symmetriesYBE}
%%%%%%%%%%%%%%%%%%%%%%%%%%%%%%%%%%%%%%%%%%%%%%%%%

The YBE (in both versions) has $N^6$ equations in $N^4$ variables when $V=\mathbb{C}^N$. Because of this, the system of equations is highly non-linear and over-determined. As $N$ grows, it becomes increasingly difficult to solve the equation using the simplest and most straightforward way of obtaining matrices. However, the symmetries of the YBE can simplify the ans\"{a}tze for the 4 by 4 matrices for small $N$, especially for $N=2$. These symmetries come in two varieties: discrete and continuous.

%%%%%%%%%%%%%%%%%%%%%%%%%%%%%%%%%%%%%%%%%%%%%%%%%
\subsubsection*{Continuous symmetries}
\label{subsubsec:Continuoussymmetries}
%%%%%%%%%%%%%%%%%%%%%%%%%%%%%%%%%%%%%%%%%%%%%%%%%

The continuous set of symmetries
\begin{equation}
    R \to \kappa (Q\otimes Q)R(Q\otimes Q)^{-1}\,,
\label{eq:Rcontinuous}
\end{equation}
are generated by the local operators $Q$ and a constant complex parameter $\kappa$. This gauge symmetry is easily verified for both the braided and non-braided variants of the YBE (\eqref{eq:YBEbraided} and \eqref{eq:YBEnonbraided}).

%%%%%%%%%%%%%%%%%%%%%%%%%%%%%%%%%%%%%%%%%%%%%%%%%
\subsubsection*{Discrete symmetries}
\label{subsubsec:Discretesymmetries}
%%%%%%%%%%%%%%%%%%%%%%%%%%%%%%%%%%%%%%%%%%%%%%%%%

To identify the discrete symmetries we write down the non-braided YBE in component form\footnote{Analogous arguments go through for the braided form as well.}:
\begin{equation}\label{eq:YBEnonbraided-component}
    R_{j_1\,j_2,\,k_1\,k_2}R_{k_1\,j_3,\,l_1\,k_3}R_{k_2\,k_3,\,l_2\,l_3}= R_{j_2\,j_3,\,k_2\,k_3} R_{j_1\,k_3,\,k_1\,l_3} R_{k_1\,k_2,\,l_1\,l_2}.
\end{equation}
For simplicity, we use the $N=2$ setting to demonstrate the concepts. In the case of an arbitrary $N$, the statements are evident. We take the following convention for the YBO:
\begin{equation}
    R=\begin{pmatrix} R_{00,\,00} & R_{00,\,01} & R_{00,\,10} & R_{00,\,11} \\
R_{01,\,00} & R_{01,\,01} & R_{01,\,10} & R_{01,\,11} \\
R_{10,\,00} & R_{10,\,01} & R_{10,\,10} & R_{10,\,11} \\
R_{11,\,00} & R_{11,01} & R_{11,\,10} & R_{11,\,11} \end{pmatrix}.
\label{eq:YBOcomponent}
\end{equation}
Then the discrete transformations\footnote{These transformations are also called the $P$, $C$ and $T$ symmetries respectively \cite{hietarinta-conferenceProceeding,hietarinta1993-JMP-Long}. The $P$ here is not to be confused with the permutation operator $P$ in \eqref{eq:permRelations}.} are given by:
\begin{eqnarray}
    & & R_{i\,j,\,k\,l}\to R_{k\,l,\,i\,j} \,,~~~\textrm{Discrete - I}  \label{eq:discrete-1} \\
& & R_{i\,j,\,k\,l} \to R_{\bar{i}\,\bar{j},\,\bar{k}\,\bar{l}} \,, ~~~\textrm{Discrete - II}\label{eq:discrete-2} \\
& & R_{i\,j,\,k\,l} \to R_{j\,i,\,l\,k}, ~~~\textrm{Discrete - III}. \label{eq:discrete-3}
\end{eqnarray}
The transformation Discrete-I \eqref{eq:discrete-1} is the matrix transpose (reflection about the main diagonal) taken in \eqref{eq:YBOcomponent} :
\begin{equation}\label{eq:RDiscrete-I}
    R^{(I)} = R^T.
\end{equation}
In the transformation Discrete-II \eqref{eq:discrete-2}, the $\bar{i}$ is the negation of $i$, i.e., $\bar{0}\equiv 1$ and $\bar{1}\equiv 0$. Two reflections are associated with this transformation: one along the primary diagonal, as in the transpose, and the other along the opposite diagonal. It makes no difference what order these reflections are made in. The resulting $R$-matrix is given by:
\begin{equation}
    R^{(II)}=\begin{pmatrix} R_{11,\,11} & R_{11,\,10} & R_{11,\,01} & R_{11,\,00} \\
R_{10,\,11} & R_{10,\,10} & R_{10,\,01} & R_{10,\,00} \\
R_{01,\,11} & R_{01,\,10} & R_{01,\,01} & R_{01,\,00} \\
R_{00,\,11} & R_{00,10} & R_{00,\,01} & R_{00,\,00} \end{pmatrix}.
\label{eq:YBOcomponent-Discrete-II}
\end{equation}
It is easily seen that this is a special case of the continuous transformation in \eqref{eq:Rcontinuous} as :
\begin{equation}\label{eq:RDiscrete-II}
    R^{(II)} = \left(X\otimes X \right)~R~\left(X\otimes X \right),
\end{equation}
where $X$ is the first Pauli matrix $X=\begin{pmatrix}
    0 & 1 \\ 1 & 0 
\end{pmatrix}$. 

The third discrete transformation \eqref{eq:discrete-3} results in the $R$-matrix :
\begin{equation}
    R^{(III)}=\begin{pmatrix} R_{00,\,00} & R_{00,\,10} & R_{00,\,01} & R_{00,\,11} \\
R_{10,\,00} & R_{10,\,10} & R_{10,\,01} & R_{10,\,11} \\
R_{01,\,00} & R_{01,\,10} & R_{01,\,01} & R_{01,\,11} \\
R_{11,\,00} & R_{11,10} & R_{11,\,01} & R_{11,\,11} \end{pmatrix}.
\label{eq:YBOcomponent-Discrete-III}
\end{equation}
This is obtained by conjugating the $R$-matrix \eqref{eq:YBOcomponent} with the permutation operator $P$ \eqref{eq:permutationC2} : 
\begin{equation}\label{eq:RDiscrete-III}
    R^{(III)} = P~R~P.
\end{equation}
These three distinct symmetries are readily confirmed to be present in both the braided \eqref{eq:YBEbraided} and the non-braided \eqref{eq:YBEnonbraided} forms of the YBE. It should be noted that the YBE is also symmetric in the presence of a combination of these discrete symmetries. Now that the continuous and discrete symmetries have been taken into consideration, we may define the equivalence class of YBE solutions.
\begin{definition}[Equivalence class of $R$-matrices]\label{def:equivalenceClass}
    Two $R$-matrices are considered equivalent if they are related by either the continuous symmetry \eqref{eq:Rcontinuous} or any of the three discrete symmetries \eqref{eq:discrete-1}-\eqref{eq:discrete-3} or any combination of the latter transformations.
\end{definition}
It should be mentioned that the three discrete transformations \eqref{eq:discrete-1}-\eqref{eq:discrete-3} are unable to create equivalence classes on their own since they do not adhere to the reflexivity property. Together, these transformations can define an equivalence class as defined in Definition \ref{def:equivalenceClass}, but only when accompanied with the continuous symmetry \eqref{eq:Rcontinuous}, which covers the reflexivity.

With this definition the constant 4 by 4 YBO's fall into 10 classes \cite{HIETARINTA-PLA} (See also Appendix of \cite{Padmanabhan_2021} for details.):
\begin{eqnarray}
    & & H3,1 = \begin{pmatrix} k & 0 & 0 & 0 \\ 0 & 0 & p & 0 \\ 0 & q & 0 & 0 \\ 0& 0 & 0 & s \end{pmatrix} ,~
H2,1=\begin{pmatrix} k^2 & 0 & 0 & 0 \\ 0 & k^2-pq & kp & 0 \\ 0 & kq & 0 & 0 \\ 0 & 0 & 0 & k^2 \end{pmatrix} , \nn\\
& & H2,2=\begin{pmatrix} k^2 & 0 & 0 & 0 \\ 0 & k^2-pq & kp & 0 \\ 0 & kq & 0 & 0 \\ 0 & 0 & 0 & -pq \end{pmatrix} ,~
H2,3=\begin{pmatrix} k & p & q & s \\ 0 & 0 & k & p \\ 0 & k & 0 & q\\ 0 & 0 & 0 & k \end{pmatrix}, \nn\\
& & H1,1=\begin{pmatrix} p^2+2pq-q^2 & 0 & 0 & p^2-q^2 \\ 0 & p^2-q^2 & p^2+q^2 & 0 \\ 0 & p^2+q^2 & p^2 -q^2 & 0 \\ p^2-q^2 & 0 & 0 & p^2-2pq-q^2 \end{pmatrix}, \nn \\
& & H1,2=\begin{pmatrix} p & 0 & 0 & k \\ 0 & p-q & p & 0 \\ 0 & q & 0 & 0 \\ 0 & 0 & 0 & -q \end{pmatrix}, ~
H1,3 =\begin{pmatrix} k^2 & -kp & kp & pq \\ 0 & 0 & k^2 & kq \\ 0 & k^2 & 0 & -kq \\ 0 & 0 & 0 & k^2 \end{pmatrix},~ 
H1,4=\begin{pmatrix} 0 & 0 & 0 & p \\ 0 & k & 0 & 0 \\ 0 & 0 & k & 0 \\ q & 0 & 0 & 0 \end{pmatrix},
\nn \\
& & H0,1=\begin{pmatrix} 1 & 0 & 0 & 1 \\ 0 & 0 & -1 & 0 \\ 0 & -1 & 0 & 0 \\ 0 & 0 & 0 & 1 \end{pmatrix}, ~
H0,2 = \begin{pmatrix} 1 & 0 & 0 & 1 \\ 0 & 1 & 1 & 0 \\ 0 & -1 & 1 & 0 \\ -1 & 0 & 0 & 1 \end{pmatrix}. 
\label{eq:Hietarinta4by4}
\end{eqnarray}
The invertible YBOs are made up of the ten classes listed here. The braided version of the constant YBE \eqref{eq:YBEbraided} is satisfied by them. In addition to these classes, the eleventh class produced by the permutation operator is also included. But since $$ \left[Q\otimes Q\right]~P~\left[Q^{-1}\otimes Q^{-1}\right] = P,$$ this is a singleton set.
The invariance of $P$ under the three discrete transformations \eqref{eq:discrete-1}-\eqref{eq:discrete-3} is also readily apparent. The corresponding non-braided YBO, which is obtained by multiplying the permutation operator, is simply the trivial identity operator if the permutation operator is assumed to be the solution of the braided YBE \eqref{eq:YBEbraided}. Conversely, the identity operator is equivalent to the permutation operator in the non-braided form when viewed as a trivial solution of the braided YBE \eqref{eq:YBEbraided}. This is in line with the fact that both of these solutions resolve the non-braided \eqref{eq:YBEnonbraided} and braided \eqref{eq:YBEbraided} YBEs. We do not need to take into account the identity and permutation operators because they are already algebraic solutions. They are trivially present in our methods since they will actually show up as generators of the algebra in the algebraic solutions we generate.

Before continuing, we check to see if the continuous transformation (gauge transformation) in \eqref{eq:Rcontinuous} can rotate the discrete actions \eqref{eq:discrete-1} and \eqref{eq:discrete-3} on each of these 10 classes into their corresponding classes. Since the eigenvalue structure of the matrices produced by the discrete transformations is identical to that of the original matrix, this is indeed to be expected. To ensure that we have the appropriate number of representatives, we must verify this\footnote{In this context, the term ``representatives" is used loosely. An equivalency class has a single representative by definition. For each of the 10 classes in \eqref{eq:Hietarinta4by4}, we mean by representatives the number of matrices of a particular class that must be compared with (using the gauge transformation \eqref{eq:Rcontinuous}) in order to determine whether a given solution belongs to that class or not.}. With the exception of the $(1,2)$ class, we discover that this is true for every class. According to our Definition \ref{def:equivalenceClass} of an equivalence class, the first \eqref{eq:discrete-1}, third \eqref{eq:discrete-3}, and a combination of these two transformations yield three matrices that are not gauge equivalent to $H1,2$ but nevertheless belong to the $(1,2)$ class:
\begin{eqnarray}\label{eq:12extra}
    H_{1,2}^{(I)}  = \left(
\begin{array}{cccc}
 p & 0 & 0 & 0 \\
 0 & p-q & q & 0 \\
 0 & p & 0 & 0 \\
 k & 0 & 0 & -q \\
\end{array}
\right)~~& ; &~~H_{1,2}^{(III)} = \left(
\begin{array}{cccc}
 p & 0 & 0 & k \\
 0 & 0 & q & 0 \\
 0 & p & p-q & 0 \\
 0 & 0 & 0 & -q \\
\end{array}
\right), \nonumber \\
H_{1,2}^{(I\times III)} & = & \left(
\begin{array}{cccc}
 p & 0 & 0 & 0 \\
 0 & 0 & p & 0 \\
 0 & q & p-q & 0 \\
 k & 0 & 0 & -q \\
\end{array}
\right).
\end{eqnarray}
These matrices have the same eigenvalues as the $H1,2$ class. However, they are not related to each other by the continuous transformations \eqref{eq:Rcontinuous}. All of this implies that the $(1,2)$ class has four representatives given by the usual $H1,2$ matrix in \eqref{eq:Hietarinta4by4} and the three additional ones given by \eqref{eq:12extra}. Thus while finding the class into which a given solution falls, we need to check the gauge equivalence to the 10 classes in \eqref{eq:Hietarinta4by4}, the permutation operator $P$ and the three matrices in \eqref{eq:12extra} that are obtained from $H1,2$ by applying the discrete transformations $I$, $III$ and $I\times III$\footnote{As noted earlier the identity matrix trivially solves the YBE and is a class of its own. The solutions we consider in this work are non-trivial and will not fall into this class.}.   

Because the second discrete transformation \eqref{eq:discrete-2} is a particular instance of the gauge transformations, as previously mentioned, we do not need to perform the aforementioned exercise. 

An essential resource for figuring out gauge equivalencies is the eigenvalues of the ten Hietarinta classes. Two operators are unquestionably not gauge equivalent if their eigenvalues differ. However, there may be a gauge transformation ($Q$ matrix) connecting the two operators if their eigenvalue structures are identical (as indicated by the multiplicity of the eigenvalues). Table \ref{tab:EigenvaluesH} contains the set of eigenvalues for each of the ten Hietarinta classes.
\begin{table}[h!]
   \centering
   \scalebox{1}{
    \begin{tabular}{|c|c|}
    \hline
    Class  & Eigenvalues \\ 
    \hline
      $H0,1$   & $\{-1,1,1,1\}$ \\ \hline
      $H0,2$   & $\{1+\mathrm{i},1+\mathrm{i},1-\mathrm{i},1-\mathrm{i}\}$\\ \hline
      $H1,1$   & $\{2 p^2, 2 p^2, -2q^2, -2q^2\}$\\ \hline
      $H1,2$   & $\{p, p, -q, -q\}$\\ \hline
      $H1,3$   & $\{-k^2, k^2, k^2, k^2\}$\\ \hline
      $H1,4$   &  $\{k,k,\pm \sqrt{pq}\}$\\ \hline
      $H2,1$   &  $\{k^2,k^2,k^2, -pq\}$\\ \hline
      $H2,2$   &  $\{k^2,k^2,-pq,-pq\}$\\ \hline 
      $H2,3$   &  $\{-k,k,k,k\}$\\ \hline
      $H3,1$   &  $\{k,s,\pm \sqrt{pq}\}$\\ 
      \hline
    \end{tabular}}
    \caption{The eigenvalues of the 10 Hietarinta classes in \eqref{eq:Hietarinta4by4}. These are the eigenvalues of the braided operators. The three matrices in \eqref{eq:12extra} have the same eigenvalues as $H1,2$.}
    \label{tab:EigenvaluesH}
\end{table}

%%%%%%%%%%%%%%%%%%%%%%%%%%%%%%%%%%%%%%%%%%%%%%%%%%%%%%%%%%%%
\section{Algebraic structures}
\label{sec:algebraicStructures}
%%%%%%%%%%%%%%%%%%%%%%%%%%%%%%%%%%%%%%%%%%%%%%%%%%%%%%%%%%%%
For the $N=2$ or qubit scenario, Hietarinta's approach to solving for the constant YBO's \cite{hietarinta-conferenceProceeding, hietarinta1993-JMP-Long} proved to be quite effective. This procedure must be repeated in order to determine the YBOs for other $N$. The computational resources that are available will inevitably limit this. To get around this, we'll solve the YBE in a representation-independent way using algebraic solutions. We will see that these solutions belong to one of the 10 Hietarinta classes \eqref{eq:Hietarinta4by4} and \eqref{eq:12extra} when the qubit or two-dimensional ($N=2$) representation is used. 

We use four algebraic structures to achieve our goals :
\begin{enumerate}
    \item A set of commuting operators.
    \item Clifford algebras.
    \item Temperley-Lieb algebras.
    \item Partition algebras.
\end{enumerate}
In this section, we write down the relations satisfied by the generators of these algebraic structures. Following this we will study how to obtain every Hietarinta class from an algebraic solution to the YBE.

%%%%%%%%%%%%%%%%%%%%%%%%%%%%%%%%%%%
\subsection{Commuting operators}
\label{subsec:commutingOP}
%%%%%%%%%%%%%%%%%%%%%%%%%%%%%%%%%%
Consider a set of commuting operators\footnote{We can also think of this as an Abelian set of operators or an Abelian algebra.}:
\begin{equation}\label{eq:Scommuting}
    \mathcal{S}_{\rm{commuting}} = \left\{A^{(\alpha)}\bigg|\left[A^{(\alpha)}, A^{(\beta)} \right]=0~;~\alpha, \beta\in \{1,2,\cdots \} \right\},
\end{equation}
with the operators $A^{(\alpha)}$ acting on the local Hilbert space $V$. The range of values that the indices $\alpha$ and $\beta$ take determines whether this collection is finite or infinite. We only need two commuting operators in this work, therefore $\alpha, \beta\in\{1,2\}$. When $V\simeq \mathbb{C}^2$, this is adequate. The reason for this is that there can only be two non-trivial 2 by 2 matrices that mutually commute. Considering this, we take:
$$ A^{(1)}\equiv A~~;~~A^{(2)}\equiv B,$$
and 
$$ \left[A,B\right] = 0.$$

%%%%%%%%%%%%%%%%%%%%%%%%%%%%%%%%%%%%%
\subsection{Anticommuting operators}
\label{subsec:anticommutingOP}
%%%%%%%%%%%%%%%%%%%%%%%%%%%%%%%%%%%%%
Consider a pair of anticommuting operators $A$ and $B$ 
\begin{equation}
    \{A, B\} = 0.
\end{equation}
Such operators can be realized using the generators of Clifford algebras $\mathbf{CL}(p,q)$ of order $p+q$, which we now define.
\begin{definition}\label{def:clifford}
    Consider a vector space with a degenerate quadratic form denoted by a multiplication. The quadratic form has a signature $p+q$ with $p$ and $q$ being positive integers. A Clifford algebra is an associative algebra generated by $p+q$ elements denoted $$\{\Gamma_1, \cdots , \Gamma_{p+q}\}.$$ They satisfy the relations
    \begin{eqnarray}
        \Gamma^2_i & = & \mathbb{1}~\textrm{for}~1\leq i\leq p, \label{eq:clifford-1} \\ 
        \Gamma^2_i & = & -\mathbb{1}~\textrm{for}~p+1\leq i\leq p+q,\label{eq:clifford-2}\\ 
        \Gamma_i\Gamma_j & = & -\Gamma_j\Gamma_i~\textrm{for}~i\neq j. \label{eq:clifford-3}
    \end{eqnarray}
\end{definition}
We use examples where both $p$ and $q$ are non-zero in the following ans\"{a}tze. We also take into account situations in which $A$ or $B$ is non-invertible but yet anticommuting. Clifford algebras can also be used to realize such operators, as we will demonstrate.

%%%%%%%%%%%%%%%%%%%%%%%%%%%%%%%%%%%%%
\subsection{Temperley-Lieb algebras}
\label{subsec:TLalgebra}
%%%%%%%%%%%%%%%%%%%%%%%%%%%%%%%%%%%%
The Temperley-Lieb (TL) algebra \cite{Temperley:1971iq}, $\mathbf{TL}_M(\eta)$ is generated by $M-1$ generators $e_i\equiv e_{i,i+1}$ with $i\in\{1,2,\cdots, M-1\}$. The integers $i$ index the local Hilbert space $V$. Each generator $e_i$ acts non-trivially on $i$, ${i+1}$, and as identity on all the other indices. They satisfy the relations
\begin{eqnarray}\label{eq:TLrelations}
    e_i^2 &=& \eta~ e_i,~~\eta\in\mathbb{C},\notag\\
    e_i e_{i\pm 1}e_i &=& e_i,\notag\\
 e_i e_j &=& e_j e_i, ~~\text{for} ~ |i-j| >1.
\end{eqnarray}
Under the operation of the permutation operator $P_i\equiv P_{i,i+1}$, these generators are typically invariant. When this is the case, they generate the Brauer algebra \cite{brauer} together with the permutation operator. The resulting algebra can then be defined as a subalgebra of the partition algebra (See Sec. \ref{subsec:partitionAlgebra}). However, the YBOs we obtain from this algebra are different from those derived from the partition algebras because the two-dimensional representation we utilize in this study does not commute with the permutation operator on $\mathbb{C}^2\otimes\mathbb{C}^2$.

%%%%%%%%%%%%%%%%%%%%%%%%%%%%%%%%%%
\subsection{Partition algebra}
\label{subsec:partitionAlgebra}
%%%%%%%%%%%%%%%%%%%%%%%%%%%%%%%%%%
The partition algebra \cite{halverson2004partitionalgebras} is defined on $M$ points, each of which has a local Hilbert space associated with it, much like the TL algebra. It is generated by three sets of operators:
\begin{eqnarray}
 & f_i~\text{for}~ i \in \{1,\dots,M\} & \nonumber \\ 
 & f_{i+\frac{1}{2}}\equiv f_{i,i+1}~\text{for}~i \in \{1,\dots,M-1\} &  
\end{eqnarray}
along with the $M-1$ permutation operators $P_i$. The operators $f_i$ act on a single space $V$ and the operators $f_{i+\frac{1}{2}}$ act on neighboring $V$'s indexed by $i$ and $i+1$. Moreover, the $f_i$ and $f_{i+\frac{1}{2}}$ operators commute among themselves. 

They satisfy the relations,
\begin{eqnarray}\label{eq:partitionrelations-1}
    f_i^2 =  f_i,~&&~f_{i+\frac{1}{2}}^2=f_{i+\frac{1}{2}}, \nonumber\\
    f_if_{i\pm\frac{1}{2}}f_i=f_i, ~&&~ f_{i\pm\frac{1}{2}}f_if_{i\pm\frac{1}{2}}=f_{i\pm \frac{1}{2}}, \nonumber \\
   f_i f_j=f_j f_i ~&&~\text{for}\; |i-j|> 1/2, 
\end{eqnarray}
The generators $f_i$ and $f_{i+\frac{1}{2}}$ are projectors like the TL generators. With the inclusion of the permutation generators $P_i$ we have another set of relations,
\begin{eqnarray}\label{eq:partitionrelations-2}
     & &P_if_if_{i+1}=f_if_{i+1}P_i=f_if_{i+1} \nonumber \\
     & &P_i f_{i+\frac{1}{2}}= f_{i+\frac{1}{2}}P_i= f_{i+\frac{1}{2}} \nonumber \\
    & &P_if_{i+j}=f_{i+j}P_i, \hspace{1cm}\text{for}\; j\neq -\frac{1}{2},0,1,\frac{3}{2}. 
\end{eqnarray}
Later on, we assume the following representation for the partition algebra generators when we assume that the local Hilbert space is $\mathbb{C}^2$.
\begin{equation}\label{eq:C2reppartition}
    f_i= \frac{\mathbb{1}+Z_i}{2}; ~~ f_{i+\frac{1}{2}}= \mathbb{1}+X_i X_{i+1}.
\end{equation}
The \eqref{eq:permutationC2} provides the permutation operator on $\mathbb{C}^2\otimes\mathbb{C}^2$. As indicated by the defining relations in \eqref{eq:partitionrelations-1}, this representation is not a pure projector, as $f_{i+\frac{1}{2}}^2=2f_{i+\frac{1}{2}}$. This has no effect on the other relations in \eqref{eq:partitionrelations-1} and \eqref{eq:partitionrelations-2}. As we shall see later, this fact has no bearing on the braid generators we build with the partition algebra generators. We might rescale the $f_{i+\frac{1}{2}}$ generator, but we ignore this alternative because it introduces a number of factors into the other relations of the partition algebra.  
We close the discussion on partition algebras with three remarks.
\begin{remark}
    This representation can easily be generalized to a qudit or $d$-dimensional setting using the clock and shift operators of $\mathbb{Z}_d$ in place of the Pauli $Z$ and $X$. This is essential while considering the higher dimensional representations of the YBO's constructed out of partition algebras.
\end{remark}
\begin{remark}
    As noted earlier in Sec. \ref{subsec:TLalgebra}, the TL algebra is a subalgebra of the partition algebra. It is easily seen that the operators
    \begin{equation}\label{eq:TlfromPartition}
        e_i = f_{i+\frac{1}{2}}f_if_{i+1}f_{i+\frac{1}{2}},
    \end{equation}
    satisfy the relations of the TL algebra \eqref{eq:TLrelations} as a consequence of the partition algebra relations \eqref{eq:partitionrelations-1}. Moreover, this realization also guarantees that the TL generators are permutation invariant. This is due to the partition algebra relations \eqref{eq:partitionrelations-2}. In this case, they satisfy the relations of the Brauer algebra \cite{brauer}.
\end{remark}
\begin{remark}
    Partition algebras and TL algebras are special cases of the more general diagram algebras. The generators of these algebras can be represented by diagrams which can be combined and manipulated using the rules of a diagrammatic calculus. This helps in understanding several of the relations satisfied by these generators \eqref{eq:partitionrelations-1}, \eqref{eq:partitionrelations-2} and \eqref{eq:TLrelations}. Some of these can be found in \cite{kauffman-knots, halverson2004partitionalgebras,Padmanabhan_2020} and we refer the interested readers to these references.
\end{remark}

%%%%%%%%%%%%%%%%%%%%%%%%%%%%%%%%%%%%%%%%%%%%%%%%%%%%%%%%%%%%
\section{Algebraic solutions for the Hietarinta classes}
\label{sec:algebraicHietarinta}
%%%%%%%%%%%%%%%%%%%%%%%%%%%%%%%%%%%%%%%%%%%%%%%%%%%%%%%%%%%%
The 10 Hietarinta classes in \eqref{eq:Hietarinta4by4} can be divided according to the number of parameters appearing in each of them \cite{hietarinta1993-JMP-Long}. The $(0,1)$ and $(0,2)$ classes have no parameters. The $(1,1)$, $(1,2)$, $(1,3)$, $(1,4)$ classes have one independent parameter up to the gauge transformations \eqref{eq:Rcontinuous}. The classes with two independent parameters are $(2,1)$, $(2,2)$ and $(2,3)$. The last class $(3,1)$ has three independent parameters. The eleventh class is the singleton set consisting of the permutation operator. We will not discuss this case as it appears in the definition of the partition algebra generators implying that this is the algebraic solution. In the rest of the section, we will discuss the different algebraic solutions that contain a given class when the $\mathbb{C}^2$ representation is chosen. We do not obtain the $(2,2)$ class by any of the algebraic ans\"{a}tze and hence we do not consider that case here. 

In each case below we compute the $Q$ matrix that maps a given algebraic solution into the proposed Hietarinta class. We adopt the following notation for the $Q$ :
\begin{equation}
    Q = \begin{pmatrix}
        q_1 & q_2 \\ 
        q_3 & q_4
    \end{pmatrix}.
\end{equation}
We search for invertible $Q$'s as required by the definition of the gauge transformations \eqref{eq:Rcontinuous} and solve for its parameters $q_i$'s, in each case.

%%%%%%%%%%%%%%%%%%%%%%%%%%%%%%%%%%
\subsection{The $H0,1$ class}
\label{subsec:H01}
%%%%%%%%%%%%%%%%%%%%%%%%%%%%%%%%%%
This class is obtained from constant YBO's constructed from Clifford algebras and partition algebras.

\paragraph{From Clifford algebras :} For a pair of anticommuting operators $A$ and $B$, the operator 
\begin{equation}\label{eq:cliffordBraid}
    R_{ij} = \alpha~A_iA_j + \beta~B_iB_j,
\end{equation}
satisfies the non-braided form of the YBE \eqref{eq:YBEnonbraided} \cite{PADMANABHAN2024116664}. Choosing
$$ A = X\left(\frac{\mathbb{1}+Z}{2}\right)=\begin{pmatrix}
    0 & 0 \\ 1 & 0
\end{pmatrix}~;~B = Z = \begin{pmatrix}
    1 & 0 \\ 0 & -1
\end{pmatrix},$$
we find the resulting operator is gauge equivalent to the $H0,1$ matrix, where 
$$Q = \begin{pmatrix}
 0 & -\frac{\sqrt{\beta } ~q_3}{\sqrt{\alpha }} \\
 q_3 & 0 \\
\end{pmatrix}, ~~ \kappa= \beta.$$
%%%%%Insert the Q operator here.
Here we have chosen the Clifford algebra $\mathbf{CL}(2,0)$ to realize the above $A$ and $B$ operators.

More generally, we can use an anticommuting pair $A$ and $B$,
$$ A = \left(
\begin{array}{cc}
-d_1 & 0 \\
c_1 & d_1 \\
\end{array}
\right),~~
B = \left(
\begin{array}{cc}
0 & 0 \\
c_2 & 0 \\
\end{array}
\right), $$
that satisfies the relations $$A^2\propto \mathbb{1};~~B^2=0.$$
Then the braided YBO rotates into $H0,1$ when
$$q_1=\frac{c_1 q_2}{2 d_1},q_3= -\frac{\sqrt{\beta} c_2 q_2}{\sqrt{\alpha} d_1},q_4= 0,\kappa = \alpha d_1^2.$$

\paragraph{From partition algebras :} Consider the YBO\footnote{We could also consider an operator $$
    \tilde{R}_{ij} = P_i - 2~f_if_{i+\frac{1}{2}}f_{i+1} - 2~f_{i+1}f_{i+\frac{1}{2}}f_i + 4~f_{i+\frac{1}{2}}f_if_{i+1}.
$$ that has properties similar to \eqref{eq:partition01-1}.} 
\begin{equation}\label{eq:partition01-1}
    \tilde{R}_{i, i+1} = P_i - 2~f_if_{i+\frac{1}{2}}f_{i+1} - 2~f_{i+1}f_{i+\frac{1}{2}}f_i + 4~f_if_{i+1}f_{i+\frac{1}{2}}.
\end{equation}
This satisfies the braided form of the YBE \eqref{eq:YBEbraided}. This is hard to check analytically. However, it is easier to analytically verify that this solution multiplied by the permutation operator satisfies the non-braided form of the YBE \eqref{eq:YBEnonbraided}. In this case, each side of the equation reduces to 
$$\mathbb{1} - 4f_if_{i+1} -4f_{i+1}f_{i+2} + 4f_if_{i+2} + 4f_if_{i+1}f_{i+\frac{1}{2}}+4f_{i+1}f_{i+2}f_{i+\frac{3}{2}}-4f_if_{i+2}f_{i,i+2}, $$
with $f_{i,i+2}=P_{i+1}f_{i+\frac{1}{2}}P_{i+1}$. The inverse of this operator is given by 
\begin{equation}
    \tilde{R}_{i, i+1}^{-1} =P_i~\left[\mathbb{1}-2\left(f_i + f_{i+1}\right) +8f_if_{i+1} - 4f_if_{i+1}f_{i+\frac{1}{2}}\right].
\end{equation}
On choosing the $\mathbb{C}^2$ representations for the partition algebra generators \eqref{eq:C2reppartition} we find that this solution is gauge equivalent to the $H0,1$ matrix {\it via}
$$Q = \begin{pmatrix}
 q_1 & 0 \\
 0 & \pm \frac{q_1}{2} \\
\end{pmatrix}, ~~ \kappa= 1.$$

%%%%%%%%%%%%%%%%%%%%%%%%%%%%%%%%%%
\subsection{The $H0,2$ class}
\label{subsec:H02}
%%%%%%%%%%%%%%%%%%%%%%%%%%%%%%%%%%
This class can be obtained from Clifford algebras.
Consider the following YBO obtained from a pair of anticommuting operators $A$ and $B$ 
\begin{eqnarray}\label{eq:11Cliffordbraid}
 \tilde{R}_{ij} = \mathbb{1} + A_iB_j.
\end{eqnarray}
These operators are realized using the order 2 Clifford algebra $\mathbf{CL}(1,1)$ with 
$$ A^2=-\mathbb{1}~~;~~B^2=\mathbb{1}.$$
In the $\mathbb{C}^2$ representation we pick the matrices
$$ A = \mathrm{i}Y=\begin{pmatrix}
    0 & 1 \\ -1 & 0 
\end{pmatrix}~~;~~B= X= \begin{pmatrix}
    0 & 1 \\ 1 & 0
\end{pmatrix}.$$ Then the above operator is precisely $H0,2$ \eqref{eq:Hietarinta4by4}.

The braid representation in \eqref{eq:11Cliffordbraid} is very similar to the braid representations constructed using Majorana fermions \cite{PhysRevLett.86.268,Kauffman_2016, Kauffman_2018}. We find that the latter also gives rise to the $(0,2)$ class as follows.
Consider a set of Majoranas $\{\gamma_1,\gamma_2, \dots ,\gamma_M\}$, generating the Clifford algebra $\mathbf{CL}(M,0)$ over the real numbers. They satisfy the relations:
\begin{equation}
    \gamma_i^2= \mathbb{1} ~ \forall i; ~~~ \gamma_i \gamma_j + \gamma_j \gamma_i=0 ~~ \text{for}~ i\neq j.
\end{equation}
The YBO satisfying the braided YBE \eqref{eq:YBEbraided} is then
\begin{equation}
    \tilde{R}_{i,i+1}= (\mathbb{1}+ \gamma_{i+1} \gamma_i).
\end{equation}
If we choose qubit representation on each site, $\gamma_i$ can be expressed in terms of a non-local string operator:
\begin{equation}
    \gamma_i = X_i \displaystyle \prod_{k=1}^{i-1} Z_k,
\end{equation}
where the $X$ and $Z$ are the first and third Pauli matrices respectively.
Though these are non-local the product $\gamma_{i+1} \gamma_i=\mathrm{i} Y_i X_{i+1}$ is local\footnote{The Majorana braid representation is non-local. Strictly, it needs to be localized \cite{Rowell_2011} before it can be compared with the solutions of Hietarinta. However, when nearest neighbor sites are selected, these non-local solutions are 4 by 4, and are therefore compared to Hietarinta's answers.}. This braid operator is precisely $H0,2$ in \eqref{eq:Hietarinta4by4}.

%%%%%%%%%%%%%%%%%%%%%%%%%%%%%%%%%%
\subsection{The $H1,1$ class}
\label{subsec:H11}
%%%%%%%%%%%%%%%%%%%%%%%%%%%%%%%%%%
This class is contained in the constant YBO's constructed using the TL algebra. In this case, the constant YBO is precisely the Jones representation \cite{Jones:1985dw} :
\begin{equation}\label{eq:braidJonesRep}
    \tilde{R}_{i, i+1}\equiv\tilde{R}_{i} = \alpha~\mathbb{1} + \beta~e_i.
\end{equation}
This satisfies the braided YBE \eqref{eq:YBEbraided} when the complex parameters $\alpha$, $\beta$ satisfy,
\begin{equation}
    \alpha^2 + \beta^2 + \eta\alpha\beta=0.
\end{equation}
To go further we require the representations of TL algebras. We use the technique demonstrated by Kulish {\it et al.} \cite{Kulish_2008} to obtain these representations. Consider a local Hilbert space $V$ with dimension $N$. Then a square matrix `$e$' of size $N^2 \times N^2$ can be constructed in terms of two $N \times N$ invertible matrices $A$ and $B$ as,
\begin{equation}\label{eq:eAB}
    e_{ab,cd}= A_{ab} B_{cd}.
\end{equation}
The projector condition $e^2 = \eta e$ yields $\eta= \text{Tr}(A^T B)$, where trace Tr is taken over $V$. The relation $e_i e_{i\pm 1}e_i = e_i$ is satisfied if and only if $A$ and $B$ satisfy the condition
\begin{equation}\label{eq:abparameter}
    (BA)^T (AB) = (AB) (BA)^T = \mathbb{1}.
\end{equation}
Then the pair
$$A =
\begin{pmatrix}
 \frac{a_2^2+\frac{1}{b_2^2}}{a_4} & a_2 \\
 a_2 & a_4 \\
\end{pmatrix}
, ~~
B= 
\begin{pmatrix}
 0 & b_2 \\
 b_2 & -\frac{2 a_2 b_2}{a_4} \\
\end{pmatrix},$$
leads to a matrix representation of $e$ for $\eta = 0$. This results in a constant $\tilde{R}$-matrix for $\beta = \rm i\alpha$,
\begin{equation}
\tilde{R} =
\begin{pmatrix}
 1 & \frac{\rm i b_2 \left(a_2^2+\frac{1}{b_2^2}\right)}{a_4} & \frac{\rm i b_2 \left(a_2^2+\frac{1}{b_2^2}\right)}{a_4} & -\frac{2 \rm i a_2 b_2 \left(a_2^2+\frac{1}{b_2^2}\right)}{a_4^2} \\
 0 & 1+\rm i a_2 b_2 & \rm i a_2 b_2 & -\frac{2 \rm i a_2^2 b_2}{a_4} \\
 0 & \rm i a_2 b_2 & 1+\rm i a_2 b_2 & -\frac{2 \rm i a_2^2 b_2}{a_4} \\
 0 & \rm i a_4 b_2 & \rm i a_4 b_2 & 1-2 \rm i a_2 b_2 \\
\end{pmatrix}.
\end{equation}
This yields just two of the operators in the $(1,1)$ class with 
$$\left\{q_2 \to -\frac{\rm i q_1 \left(a_2 b_2-1\right)}{\sqrt{\rm i \left(a_2 b_2+1\right){}^2}},~ q_3 \to \frac{a_4 b_2 q_1}{a_2 b_2+1}, ~q_4\to -\frac{\rm i a_4 b_2 q_1}{\sqrt{\rm i \left(a_2 b_2+1\right){}^2}},~p\to \frac{1}{\sqrt{2} \sqrt{\kappa}},~q\to \frac{\rm i}{\sqrt{2} \sqrt{\kappa}}\right\}.$$

%%%%%%%%%%%%%%%%%%%%%%%%%%%%%%%%%%
\subsection{The $H1,2$ class}
\label{subsec:H12}
%%%%%%%%%%%%%%%%%%%%%%%%%%%%%%%%%%
The operators in this class can only be obtained from the algebraic solutions constructed using the partition algebras. By our Definition \ref{def:equivalenceClass} of an equivalence class of solutions we recall that there are four representatives for this class, given by $H1,2$ in \eqref{eq:Hietarinta4by4} and by the three matrices $H_{1,2}^{(I)}$, $H_{1,2}^{(III)}$, $H_{1,2}^{(I\times III)}$ in \eqref{eq:12extra}. The latter is obtained from $H1,2$ by the action of the first, third, and a product of the two discrete transformations respectively. We now present the algebraic solutions for each of these four representatives of the $(1,2)$ Hietarinta class.

We require four algebraic solutions that will map to the four representatives of the $(1,2)$ class under the gauge transformations \eqref{eq:Rcontinuous}. These are obtained from the following ansatz. Consider the braided YBO 
\begin{equation}
     \tilde{R}_{i,i+1} = \lambda~ \mathbb{1}+ \mu ~f_{i+\frac{1}{2}} f_i + \nu~ f_{i+\frac{1}{2}} f_{i+1}.
\end{equation}
Substituting this into the braided YBE \eqref{eq:YBEbraided} we find
\begin{align}
    &\tilde{R}_{i,i+1}\tilde{R}_{i+1,i+2}\tilde{R}_{i,i+1} -\tilde{R}_{i+1,i+2}\tilde{R}_{i,i+1}\tilde{R}_{i+1,i+2} \nonumber\\
   & =(\lambda^2 \mu+ \lambda \mu^2 + \nu \mu^2+ \lambda \mu \nu) (f_{i+\frac{1}{2}} f_i -f_{i+\frac{3}{2}} f_{i+1}) \notag\\
    & ~~~~~~~~~~~~~~~~~~~~~~~ + (\lambda^2 \nu+ \lambda \nu^2 + \mu \nu^2+ \lambda \mu \nu)  (f_{i+\frac{1}{2}} f_{i+1}-f_{i+\frac{3}{2}} f_{i+2}).
\end{align}
The above equation will be zero if and only if 
$$\lambda^2 \mu+ \lambda \mu^2 + \nu \mu^2+ \lambda \mu \nu =0, ~\text{or}~~
\lambda^2 \nu+ \lambda \nu^2 + \mu \nu^2+ \lambda \mu \nu =0.$$
Both conditions simplify to 
\begin{equation}
    \lambda^2 + (\mu +\nu)\lambda+ \mu \nu = 0
    \implies \lambda= -\mu, ~\text{or}~~ -\nu.
\end{equation}
This gives us two of the four solutions. To obtain the remaining two solutions we use the ansatz
\begin{equation}
     \tilde{R}_{i,i+1} = \lambda~ \mathbb{1}+ \mu ~f_i f_{i+\frac{1}{2}} + \nu~  f_{i+1} f_{i+\frac{1}{2}},
\end{equation}
to determine the braided YBO's. We find that 
\begin{align}
    &\tilde{R}_{i,i+1}\tilde{R}_{i+1,i+2}\tilde{R}_{i,i+1} -\tilde{R}_{i+1,i+2}\tilde{R}_{i,i+1}\tilde{R}_{i+1,i+2} \nonumber\\
   & =(\lambda^2 \mu+ \lambda \mu^2 + \nu \mu^2+ \lambda \mu \nu) (f_i f_{i+\frac{1}{2}} -f_{i+1} f_{i+\frac{3}{2}} ) \notag\\
    & ~~~~~~~~~~~~~~~~~~~~~~~ + (\lambda^2 \nu+ \lambda \nu^2 + \mu \nu^2+ \lambda \mu \nu)  (f_{i+1} f_{i+\frac{1}{2}}- f_{i+2} f_{i+\frac{3}{2}}),
\end{align}
is satisfied when
$$\lambda= -\mu, ~\text{or}~~ -\nu.$$
These are the remaining two solutions. Next we look at the precise mappings to the $(1,2)$ class. In each case we will write down the algebraic inverses as well.

%%%%%%%%%%%%%%%%%%%%%%%%%%%
\paragraph{For $H1,2$ :}
%%%%%%%%%%%%%%%%%%%%%%%%%%%
Consider the YBO that satisfies the braided YBE
\begin{equation}\label{eq:partition12-1}
    \tilde{R}_{i,i+1} = P_i - 2f_{i+\frac{1}{2}}f_if_{i+1}.
\end{equation}
When multiplied by the permutation operator $P_i$, it reduces both sides of the non-braided YBE \eqref{eq:YBEnonbraided} to
$$\mathbb{1} - 2f_{i+\frac{1}{2}}f_if_{i+1} - 2P_{i+1}f_{i+\frac{1}{2}}P_{i+1}f_if_{i+2} - 2f_{i+\frac{3}{2}}f_{i+1}f_{i+2} + 4P_{i+1}f_{i+\frac{1}{2}}P_{i+1}f_if_{i+1}f_{i+2}.$$
It is interesting to note that this operator is its own inverse. This is verified from the relations of the partition algebra \eqref{eq:partitionrelations-1} and \eqref{eq:partitionrelations-2}. As it also satisfies the braid relation \eqref{eq:YBEbraided} we tend to conclude that it is just a realization of the permutation group. However this operator is not hermitian unlike the permutation operator.

The operator \eqref{eq:partition12-1} is gauge equivalent to $H1,2$, when the representation \eqref{eq:C2reppartition} is chosen, by the transformation
$$\left\{q_1\to 0,~ q_4\to 0,~ k\to -\frac{2 q_2^2}{\kappa q_3^2},~ p\to \frac{1}{\kappa},~ q\to \frac{1}{\kappa}\right\}.$$

The other braided YBO that is gauge equivalent to $H1,2$, when the representation \eqref{eq:C2reppartition} is chosen, is given by
\begin{equation}\label{eq:partition12-2}
    \tilde{R}_{i,i+1} = \alpha~\left(\mathbb{1} - f_{i+1}f_{i+\frac{1}{2}}\right) + \beta~f_if_{i+\frac{1}{2}}~;~\alpha,\beta\in\mathbb{C}.
\end{equation}
The inverse is given by 
\begin{equation}
    \tilde{R}_{i,i+1}^{-1} = \frac{1}{\alpha}~\left(\mathbb{1} - f_{i}f_{i+\frac{1}{2}}\right) + \frac{1}{\beta}~f_{i+1}f_{i+\frac{1}{2}}~;~\alpha,\beta\in\mathbb{C}.
\end{equation}
The $Q$ operator in this case is given by
$$\left\{q_2\to 0,~ q_3\to 0,~ k\to \frac{q_4^2 \left(\beta -\alpha \right)}{\kappa q_1^2},~ p\to \frac{\beta }{\kappa},~ q\to -\frac{\alpha }{\kappa}\right\}.$$

%%%%%%%%%%%%%%%%%%%%%%%%%%%%%%%%%%%%
\paragraph{For $H_{1,2}^{(I)}$ :}
%%%%%%%%%%%%%%%%%%%%%%%%%%%%%%%%%%%%
The YBO that satisfies the braided YBE
\begin{equation}\label{eq:partition12I-1}
    \tilde{R}_{i,i+1} = P_i - 2f_if_{i+1}f_{i+\frac{1}{2}}.
\end{equation}
This simplifies both sides of the non-braided YBE to
$$\mathbb{1} -2f_if_{i+1}f_{i+\frac{1}{2}} -2f_{i+1}f_{i+2}f_{i+\frac{3}{2}} -2f_if_{i+2}P_{i+1}f_{i+\frac{1}{2}}P_{i+1} + 4f_if_{i+1}f_{i+2}P_{i+1}f_{i+\frac{1}{2}}P_{i+1}.$$
Notice that this operator is the adjoint of the operator \eqref{eq:partition12-1}. It also squares to the identity and resembles a non-hermitian permutation-like operator.

This operator is gauge equivalent to $H_{1,2}^{(I)}$ \eqref{eq:12extra}, when the representation \eqref{eq:C2reppartition} is chosen, by the transformation
$$\left\{q_1\to 0,~ q_4\to 0,~ k\to -\frac{2 q_3^2}{\kappa q_2^2},~ p\to \frac{1}{\kappa},~ q\to \frac{1}{\kappa}\right\}.$$

In this case, the other YBO is given by
\begin{equation}\label{eq:partition12I-2}
    \tilde{R}_{i,i+1} = \alpha~\left(\mathbb{1} - f_{i+\frac{1}{2}}f_{i+1}\right) + \beta~f_{i+\frac{1}{2}}f_i~;~\alpha,\beta\in\mathbb{C}.
\end{equation}
Its inverse is
\begin{equation}
    \tilde{R}_{i, i+1}^{-1} = \frac{1}{\alpha}~\left(\mathbb{1} - f_{i+\frac{1}{2}}f_{i}\right) + \frac{1}{\beta}~f_{i+\frac{1}{2}}f_{i+1}~;~\alpha,\beta\in\mathbb{C}.
\end{equation}
This is gauge equivalent to $H_{1,2}^{(I)}$ \eqref{eq:12extra}, when the representation \eqref{eq:C2reppartition} is chosen, by the transformation
$$\left\{q_2\to 0,~ q_3\to 0,~ k\to \frac{q_1^2 \left(\beta -\alpha \right)}{\kappa q_4^2},~ p\to \frac{\beta }{\kappa},~ q\to -\frac{\alpha }{\kappa}\right\}.$$

%%%%%%%%%%%%%%%%%%%%%%%%%%%%%%%%%%%%
\paragraph{For $H_{1,2}^{(III)}$ :}
%%%%%%%%%%%%%%%%%%%%%%%%%%%%%%%%%%%%
Now the braided YBO is given by
\begin{equation}\label{eq:partition12III}
    \tilde{R}_{i, i+1} = \alpha~\left(\mathbb{1} - f_{i}f_{i+\frac{1}{2}}\right) + \beta~f_{i+1}f_{i+\frac{1}{2}}~;~\alpha,\beta\in\mathbb{C}.
\end{equation}
The inverse of this operator
\begin{equation}
    \tilde{R}_{i,i+1}^{-1} = \frac{1}{\alpha}~\left(\mathbb{1} - f_{i+1}f_{i+\frac{1}{2}}\right) + \frac{1}{\beta}~f_if_{i+\frac{1}{2}}~;~\alpha,\beta\in\mathbb{C}.
\end{equation}
The $Q$ operator is given by 
$$\left\{q_2\to 0,~ q_3\to 0,~ k\to \frac{q_4^2 \left(\beta -\alpha \right)}{\kappa q_1^2},~ p\to \frac{\beta }{\kappa},~ q\to -\frac{\alpha }{\kappa}\right\}.$$

%%%%%%%%%%%%%%%%%%%%%%%%%%%%%%%%%%%%%%%%%%%%%%
\paragraph{For $H_{1,2}^{(I\times III)}$ :}
%%%%%%%%%%%%%%%%%%%%%%%%%%%%%%%%%%%%%%%%%%%%%%
The last representative of the $(1,2)$ class is obtained from the qubit representation of the braided YBO,
\begin{equation}\label{eq:partition12IxIII}
    \tilde{R}_{i,i+1} = \alpha~\left(\mathbb{1} - f_{i+\frac{1}{2}}f_{i}\right) + \beta~f_{i+\frac{1}{2}}f_{i+1}~;~\alpha,\beta\in\mathbb{C}.
\end{equation}
The inverse of this operator
\begin{equation}
    \tilde{R}_{i,i+1}^{-1} = \frac{1}{\alpha}~\left(\mathbb{1} - f_{i+\frac{1}{2}}f_{i+1}\right) + \frac{1}{\beta}~f_{i+\frac{1}{2}}f_i~;~\alpha,\beta\in\mathbb{C}.
\end{equation}
The $Q$ operator is given by 
$$\left\{q_2\to 0,~ q_3\to 0,~ k\to \frac{q_1^2 \left(\beta -\alpha \right)}{\kappa q_4^2},~ p\to \frac{\beta }{\kappa},~ q\to -\frac{\alpha }{\kappa}\right\}.$$

\begin{remark}
    The first and third discrete transformations (see \eqref{eq:discrete-1} and \eqref{eq:discrete-3}) also relate the algebraic expressions for the four braided YBOs \eqref{eq:partition12-2}, \eqref{eq:partition12I-2}, \eqref{eq:partition12III}, and \eqref{eq:partition12IxIII}. Given that the partition algebra generators are projectors and hence invariant under the transpose operation, this should be the case. Accordingly, we do not obtain the counterparts of the two braided YBOs in \eqref{eq:partition12-1} and \eqref{eq:partition12I-1} in $H_{1,2}^{(III)}$ and $H_{1,2}^{(I\times III)}$ as they remain unchanged under conjugation via the permutation operation. 
\end{remark}

%%%%%%%%%%%%%%%%%%%%%%%%%%%%%%%%%%
\subsection{The $H1,3$ class}
\label{subsec:H13}
%%%%%%%%%%%%%%%%%%%%%%%%%%%%%%%%%%
This class is also obtained from the Jones representation \eqref{eq:braidJonesRep} when the $A$ and $B$ matrices take the form 
$$A= 
\begin{pmatrix}
 a_1 & a_2 \\
 -a_2 & 0 \\
\end{pmatrix},~
B=
\begin{pmatrix}
 0 & \frac{1}{a_2} \\
 -\frac{1}{a_2} & b_4 \\
\end{pmatrix}.$$
The above pair leads to the matrix representation of $e$ \eqref{eq:eAB} for $\eta = 2$. The corresponding constant YBO for $\beta = -\alpha$ is 
\begin{equation}
\tilde{R}=
\begin{pmatrix}
 1 & -\frac{a_1}{a_2} & \frac{a_1}{a_2} & -a_1 b_4 \\
 0 & 0 & 1 & -a_2 b_4 \\
 0 & 1 & 0 & a_2 b_4 \\
 0 & 0 & 0 & 1 \\
\end{pmatrix}.
\end{equation}
This falls into $H1,3$ class with 
$$\left\{q_3 \to 0,~ k\to \frac{1}{\sqrt{\kappa}},~ p\to \frac{a_1 q_4 }{\sqrt{\kappa} a_2  q_1},~ q\to -\frac{a_2 b_4 q_4}{\sqrt{\kappa} q_1}\right\}.$$

%%%%%%%%%%%%%%%%%%%%%%%%%%%%%%%%%%
\subsection{The $H1,4$ class}
\label{subsec:H14}
%%%%%%%%%%%%%%%%%%%%%%%%%%%%%%%%%%
As in the $(0,1)$ class we have two ways to obtain the $(1,4)$ class, from Clifford algebras and from TL algebras {\it via} the Jones representation.

\paragraph{From Clifford algebras :} We use the constant YBO \eqref{eq:cliffordBraid} with $A$ and $B$ operators given by
$$A=
\begin{pmatrix}
0 & 1 \\
-1 & 0 \\
\end{pmatrix}~~;~~
B = 
\begin{pmatrix}
1 & 0 \\
0 & -1 \\
\end{pmatrix}.$$
These satisfy $$A^2 = -\mathbb{1};~~B^2 = \mathbb{1},$$ and hence can be realized using Clifford algebras of order 2, $\mathbf{CL}(1,1)$.
The resulting YBO rotates to $H_{1,4}$ under the gauge transformation
$$\left\{\kappa\to -\frac{2 \alpha }{k-\sqrt{p} \sqrt{q}},q_1\to -q_2,q_3\to -\frac{\sqrt[4]{q} q_2}{\sqrt[4]{p}},q_4\to -\frac{\sqrt[4]{q} q_2}{\sqrt[4]{p}},\beta \to \frac{\alpha  \left(k+\sqrt{p} \sqrt{q}\right)}{\sqrt{p} \sqrt{q}-k}\right\}.$$

\paragraph{From TL algebras :} Choose the $A$ and $B$ operators as 
$$A=
\begin{pmatrix}
 \frac{a_2^2}{a_4}+\frac{1}{b_1} &  a_2 \\
 a_2 & a_4 \\
\end{pmatrix},~ 
B=
\begin{pmatrix}
 b_1 & -\frac{a_2 b_1}{a_4} \\
 -\frac{a_2 b_1}{a_4} & \frac{b_1 a_2^2+a_4}{a_4^2} \\
\end{pmatrix}.$$
This pair leads to the matrix representation of $e$ \eqref{eq:eAB} for $\eta = 2$, resulting in the constant YBO for $\beta = -\alpha$ \eqref{eq:braidJonesRep},
\begin{equation}
\tilde{R}=
\begin{pmatrix}
 -\frac{a_2^2 b_1}{a_4} & \frac{a_2 \left(a_2^2 b_1+a_4\right)}{a_4^2} & \frac{a_2 \left(a_2^2 b_1+a_4\right)}{a_4^2} & -\frac{\left(a_2^2 b_1+a_4\right){}^2}{a_4^3 b_1} \\
 -a_2 b_1 & \frac{a_2^2 b_1}{a_4}+1 & \frac{a_2^2 b_1}{a_4} & -\frac{a_2 \left(a_2^2 b_1+a_4\right)}{a_4^2} \\
 -a_2 b_1 & \frac{a_2^2 b_1}{a_4} & \frac{a_2^2 b_1}{a_4}+1 & -\frac{a_2 \left(a_2^2 b_1+a_4\right)}{a_4^2} \\
 -a_4 b_1 & a_2 b_1 & a_2 b_1 & -\frac{a_2^2 b_1}{a_4} \\
\end{pmatrix}.
\end{equation}
This falls into $H_{1,4}$ class with 
$$\left\{q_3\to 0,~ q_4\to \frac{a_4 q_2}{a_2},~k\to \frac{1}{\kappa},~p\to -\frac{a_4 q_2^2}{\kappa a_2^2 b_1  q_1^2},~q\to -\frac{a_2^2 b_1 q_1^2}{\kappa a_4  q_2^2}\right\}.$$

%%%%%%%%%%%%%%%%%%%%%%%%%%%%%%%%%%
\subsection{The $H2,1$ class}
\label{subsec:H21}
%%%%%%%%%%%%%%%%%%%%%%%%%%%%%%%%%%
This class is also obtained from the TL algebras {\it via} the Jones representation. For this case we choose the $A$ and $B$ matrices as 
$$A= 
\begin{pmatrix}
 0 & a_2 \\
 -a_2 & 0 \\
\end{pmatrix},~ 
B =
\begin{pmatrix}
 0 & b_2 \\
 -\frac{1}{a_2^2 b_2} & b_4 \\
\end{pmatrix}.$$
The above pair leads to the matrix representation of $e$ \eqref{eq:eAB} for $\eta = a_2 b_2+\frac{1}{a_2 b_2}$, that results in the constant YBO for $\beta = -\alpha  a_2 b_2$,
\begin{equation}
\tilde{R}=
\begin{pmatrix}
 1 & 0 & 0 & 0 \\
 0 & 1-a_2^2 b_2^2 & 1 & -a_2^2 b_2 b_4 \\
 0 & a_2^2 b_2^2 & 0 & a_2^2 b_2 b_4 \\
 0 & 0 & 0 & 1 \\
\end{pmatrix}.
\end{equation}
This falls into a one-parameter sub class of the $(2,1)$ class with 
$$\left\{q_3\to 0,~q_4 \to \frac{q_2 \left(\frac{1}{a_2^2}-b_2^2\right)}{b_2 b_4},~p\to \frac{1}{\sqrt{\kappa}},~q\to a_2^2 b_2^2 \frac{1}{\sqrt{\kappa}},~ k\to \frac{1}{\sqrt{\kappa}}\right\}.$$

%%%%%%%%%%%%%%%%%%%%%%%%%%%%%%%%%%
\subsection{The $H2,3$ class}
\label{subsec:H23}
%%%%%%%%%%%%%%%%%%%%%%%%%%%%%%%%%%
This class is obtained from a pair of operators that are either both projectors or are both nilpotent. Furthermore, they satisfy $AB=BA=0$ and so can be considered as both commuting and anticommuting. We will use the algebraic solutions constructed from commuting operators to find the YBO's.

Consider a pair of commuting projectors: 
\begin{equation}
    A=\left(
\begin{array}{cc}
a_1 & b_1 \\
0 & 0 \\
\end{array}
\right),~~
B = \left(
\begin{array}{cc}
0 & b_2 \\
0 & -\frac{b_2 a_1}{b_1} \\
\end{array}
\right),
\end{equation}
with the matrix entries being complex parameters.
It is easily seen that they satisfy the properties
$$A^2=A;~~B^2=B;~~A.B=0.$$
Substituting them into the ansatz \eqref{eq:YBOcommuting}, 
we can get $H2,3$\footnote{Note that for $H2,3$, we have solved for the internal parameters of the operators $A, B, Q, \text{and}~ \kappa$.}
$$\{q_3= 0,~a_1= 0,\beta _1= \frac{q q_4-\alpha _1 b_1 k q_1}{b_2 k q_1},~\beta _2= \frac{p q_4-\alpha _2 b_1 k q_1}{b_2 k q_1},~\gamma _1= \frac{q_4^2 s-b_2^2 \gamma _4 k q_1^2}{b_1^2 k q_1^2},~\kappa=\frac{1}{k}\}.$$

Next, we consider a pair of commuting nilpotent operators 
\begin{equation}
    A= X \frac{(\mathbb{1}+Z)}{2}, ~~ B=Y \frac{(\mathbb{1}+Z)}{2}.
\end{equation}
Substituting them into the algebraic solution \eqref{eq:YBOcommuting}, we obtain the $H2,3$ class with the following gauge conditions
\begin{equation}
    \left\{q_1\to 0,~k\to \frac{1}{n},~p\to \frac{q_2 (\alpha_2 +i \beta_2 )}{\kappa q_3},~q\to \frac{q_2 (\alpha_1 +i \beta_1 )}{\kappa q_3},~ s\to \frac{q_2^2 (\gamma_1 -\gamma_4 +i \gamma_3 +i \gamma_2 )}{\kappa q_3^2}\right\}.
\end{equation}

%%%%%%%%%%%%%%%%%%%%%%%%%%%%%%%%%%
\subsection{The $H3,1$ class}
\label{subsec:H31}
%%%%%%%%%%%%%%%%%%%%%%%%%%%%%%%%%%
There are three ways to arrive at the $(3,1)$ class through algebraic solutions. These include the solutions constructed from commuting operators, from anticommuting operators realized using Clifford algebras, and from partition algebras. 

\paragraph{From commuting operators :} For a pair of commuting operators $A$ and $B$ the operator
\begin{eqnarray}
    R_{ij} & = & \mathbb{1} + \alpha_1~A_i+\alpha_2~A_j + \beta_1~B_i + \beta_2~B_j \nonumber \\
    & + & \gamma_1~A_iA_j + \gamma_2~A_iB_j + \gamma_3~B_iA_j + \gamma_4~B_iB_j, \label{eq:YBOcommuting}
\end{eqnarray}
with complex parameters $\alpha$'s, $\beta$'s and $\gamma$'s, readily satisfies the non-braided form of the YBE \eqref{eq:YBEnonbraided}. This follows from the index structure of the non-braided YBE \eqref{eq:YBEnonbraided}. In fact, this operator solves the non-braided YBE when the $R$-matrices on each side of the equation are written in any order. However, note that the YBO's of \eqref{eq:Scommuting} multiplied by the permutation operator, solve the braided form of the YBE \eqref{eq:YBEbraided} in a more non-obvious manner. At first glance, these solutions may appear to be trivial solutions as everything commutes with each other. Nevertheless, we will now see that it contains the $(3,1)$ Hietarinta class. 

Consider the following $\mathbb{C}^2$ representations of the pair $A$, $B$ :
\begin{equation}
    A= \left(
\begin{array}{cc}
a_1 & 0 \\
0 & d_1 \\
\end{array}
\right),~~
B = \left(
\begin{array}{cc}
a_2 & 0 \\
0 & d_2 \\
\end{array}
\right),
\end{equation}
with $a_i, d_i\in\mathbb{C}$.
Substituting this into \eqref{eq:YBOcommuting} results in a diagonal YBO ($R$-matrix) satisfying the non-braided form of the YBE \eqref{eq:YBEnonbraided}. The braided form, obtained by multiplying the latter with the permutation operator, transforms into $H3,1$ when
\begin{eqnarray*}
 q_1&=& 0,~q_4= 0,\\p&=& \frac{\alpha _0+\alpha _2 a_1+\alpha _7 a_1 d_1+\alpha _5 a_1 d_2+\alpha _4 a_2+\alpha _6 a_2 d_1+\alpha _8 a_2 d_2+\alpha _1 d_1+\alpha _3 d_2}{\kappa},\\q &=& \frac{\alpha _0+\alpha _1 a_1+\alpha _7 a_1 d_1+\alpha _6 a_1 d_2+\alpha _3 a_2+\alpha _5 a_2 d_1+\alpha _8 a_2 d_2+\alpha _2 d_1+\alpha _4 d_2}{\kappa},\\s&=& \frac{\alpha _0+\alpha _7 a_1^2+\alpha _1 a_1+\alpha _2 a_1+\alpha _5 a_1 a_2+\alpha _6 a_1 a_2+\alpha _8 a_2^2+\alpha _3 a_2+\alpha _4 a_2}{\kappa},\\k&=& \frac{\alpha _0+\alpha _7 d_1^2+\alpha _1 d_1+\alpha _2 d_1+\alpha _5 d_1 d_2+\alpha _6 d_1 d_2+\alpha _8 d_2^2+\alpha _3 d_2+\alpha _4 d_2}{\kappa}.
\end{eqnarray*}

\paragraph{From Clifford algebras :} Consider a pair of anticommuting projectors represented on $\mathbb{C}^2$,
\begin{equation}
    A=\left(
\begin{array}{cc}
a_1 & b_1 \\
0 & 0 \\
\end{array}
\right),~~
B = \left(
\begin{array}{cc}
0 & b_2 \\
0 & -\frac{b_2 a_1}{b_1} \\
\end{array}
\right),
\end{equation}
with the matrix entries being complex parameters. These operators satisfy
$$A^2=A;~~B^2=B;~~A.B=0.$$
Thus these operators also commute with each other. Hence we can substitute them in the YBO \eqref{eq:YBOcommuting} to obtain the YBO,
\begin{equation}
    R_{ij} = \alpha A_i A_j + \beta B_i B_j + \gamma A_i B_j + \delta B_i A_j.
\end{equation}
Then the braided matrix \(\check{R} = P \cdot R_{ij}\) is identified with 
$H3,1$ if the following conditions hold: 
\begin{eqnarray*}
   q_1= 0,q_4= \frac{b_1 q_3}{a_1},&& p= -\frac{a_1^2 b_2 \gamma }{b_1 \kappa},\\ q= -\frac{a_1^2 b_2 \delta }{b_1 \kappa},&&~s= \frac{\alpha  a_1^2}{\kappa},~ k= \frac{a_1^2 \beta  b_2^2}{b_1^2 \kappa}.
\end{eqnarray*}

\paragraph{From partition algebras :} Consider the following YBO's satisfying the braided form of the YBE \eqref{eq:YBEbraided} \cite{Padmanabhan_2020}, 
\begin{eqnarray}
    \tilde{R}_{i, i+1} & = & P_i~\left(\mathbb{1} + \alpha~f_i + \beta~f_{i+1} + \gamma~f_if_{i+1} \right), \label{eq:partition31-1}\\
    \tilde{R}_{i, i+1} & = & P_i~\left(\mathbb{1} + \alpha~f_{i+\frac{1}{2}} \right), \label{eq:partition31-2}
\end{eqnarray}
with the complex parameters $\alpha$, $\beta$ and $\gamma$. It is easily verified using the partition algebra relations \eqref{eq:partitionrelations-1} and \eqref{eq:partitionrelations-2} that these operators satisfy the braided YBE \eqref{eq:YBEbraided}. In fact, it is even easier to check that they satisfy the non-braided YBE upon multiplication by the permutation operator $P$. We then observe that both these solutions are special cases of the ones obtained from commuting operators. For the partition algebra representations on $\mathbb{C}^2$ \eqref{eq:C2reppartition}, the solution in \eqref{eq:partition31-1} is the case when both $A$ and $B$ are the same diagonal matrix and the solution in \eqref{eq:partition31-2} is the case when both $A=B=X$. Thus it is clear these two solutions fall into the $(3,1)$ Hietarinta class.

Closely related to the partition algebra solution \eqref{eq:partition31-1} is another algebraic solution,
\begin{equation}\label{eq:partition31-3}
    \tilde{R}_{i, i+1} = P_i + \alpha~f_if_{i+\frac{1}{2}}f_{i+1} + \beta~f_{i+1}f_{i+\frac{1}{2}}f_i~;~\alpha, \beta\in\mathbb{C}.
\end{equation}
This satisfies the braided YBE \eqref{eq:YBEbraided}. Though it looks rather non-trivial in the braided form, its non-braided form, obtained by multiplying by $P_i$, reduces to a YBO that is a linear combination of just the identity operator $\mathbb{1}$, $f_i$ and $f_{i+1}$ due to the partition algebra relations \eqref{eq:partitionrelations-1} and \eqref{eq:partitionrelations-2}. This provides an example of two solutions that look entirely different in the braided form even though their non-braided forms are quite similar.

%%%%%%%%%%%%%%%%%%%%%%%%%%%%%%%%%%
\subsection{Summary}
\label{subsec:summary}
%%%%%%%%%%%%%%%%%%%%%%%%%%%%%%%%%%
We obtain nine of the ten Hietarinta classes in \eqref{eq:Hietarinta4by4}. While some of these nine classes are contained in algebraic solutions with distinct sources, others can be acquired from just one of the algebraic solutions. The main content of this work is captured in Table \ref{tab:summary}, which provides a clear overview of these.

\begin{table}[h!]
   \centering
\begin{tabular}{ |c|c|c|c|c| }
 \hline
Class &  Clifford & Commuting & Temperley-Lieb & Partition algebra \\
\hline \hline
$H_{3,1}$ & \checkmark & \checkmark &  & \checkmark \\
\hline
$H_{2,1}$ & &  & \checkmark - &  \\
\hline
$H_{2,2}$ & &  &  & \\
\hline 
$H_{2,3}$ & \checkmark & \checkmark & &  \\
\hline
$H_{1,1}$ & &  & \checkmark - &   \\
\hline
$H_{1,2}$ & &  & & \checkmark \\
\hline
$H_{1,3}$ & &  & \checkmark & \\
\hline
$H_{1,4}$ & \checkmark &  & \checkmark &  \\
\hline
$H_{0,1}$ & \checkmark &  & & \checkmark \\
\hline
$H_{0,2}$ & \checkmark &  & &  \\
\hline
\end{tabular}
\caption{Summary of results indicating all possible methods to generate a given Hietarinta class. The cases $\checkmark -$ indicate algebraic solutions that fall into a subclass of the specified class. }
\label{tab:summary}
\end{table}

%%%%%%%%%%%%%%%%%%%%%%%%%%%%%%%%%%%%%%%%%%%%%%%%%
\section{Conclusion}
\label{sec:conclusion}
%%%%%%%%%%%%%%%%%%%%%%%%%%%%%%%%%%%%%%%%%%%%%%%%%
In this paper, we have demonstrated various algebraic techniques for creating braid group generators, or constant YBOs. We can create braid generators in any dimension owing to the representation-independent solutions that are developed. Specifically, our solutions are gauge equivalent to Hietarinta's constant 4 by 4 solutions when the local Hilbert space is two dimensional (qubit representation) \cite{HIETARINTA-PLA,hietarinta1993-JMP-Long,Hietarinta1993-BookChapter,hietarinta-conferenceProceeding}. Ten of the eleven Hietarinta classes, including the singleton that makes up the permutation group generator, are recovered. Table \ref{tab:summary} provides a summary of the various algebraic solutions that give rise to these classes. 

During this process, we have found a number of novel algebraic solutions to the braid relation, also known as the constant Yang-Baxter equation, that may be compared to the Jones representation. We derive our solutions from Clifford algebras and partition algebras, which are generalizations of the Temperley-Lieb algebra. These are :
\begin{enumerate}
    \item Equation \eqref{eq:partition01-1} generating the $(0,1)$ class.
    \item Equation \eqref{eq:11Cliffordbraid} generating the $(0,2)$ class.
    \item Equations \eqref{eq:partition12-1} and \eqref{eq:partition12-2} generating the subclass of $(1,2)$ from $H1,2$. 
    \item Equations \eqref{eq:partition12I-1} and \eqref{eq:partition12I-2} generating the subclass of $(1,2)$ from $H_{1,2}^{(I)}$.
    \item Equation \eqref{eq:partition12III} generating the subclass of $(1,2)$ from $H_{1,2}^{(III)}$.
    \item Equation \eqref{eq:partition12IxIII} generating the subclass of $(1,2)$ from $H_{1,2}^{(I\times III)}$.
    \item Equation \eqref{eq:partition31-3} generating the $(3,1)$ class.
\end{enumerate}
The $(2,2)$ class has remained elusive to our methods. However, we do find this in a representation dependent manner.  For example, the following operator 
\begin{equation}
    R_{i, i+1} = -\frac{1}{2}\left(\lambda_1+\lambda_2\right)~P_i + \lambda_1~P_if_i + \lambda_2~f_iP_i - \frac{\left(\lambda_1^2+\lambda_2^2\right)}{\lambda_1+\lambda_2}~f_{i+1},
\end{equation}
constructed from partition algebra generators is a solution of the braided YBE for the representation \eqref{eq:C2reppartition}. This solution falls into the $(2,2)$ class. This solution is not an algebraic solution of the YBE. There are more ways of finding this class from ans\"{a}tze constructed using partition algebras. But all of them are representation-dependent.

There are a few directions that this work initiates :
\begin{enumerate}
    \item Building 9 by 9 constant YBOs that act on local Hilbert spaces of dimension 3 is an immediate use of the algebraic solutions in this study. Consider the braided YBO's three-dimensional representation of the $(1,2)$ class, \eqref{eq:partition12-2}, as an example,
    \begin{equation}\label{eq:partition-9by9}
 \tilde{R}_{ij} = \begin{pmatrix}
 \beta  & 0 & 0 & 0 & 0 & \beta -\alpha  & 0 & \beta -\alpha  & 0 \\
 0 & \alpha +\beta  & 0 & \beta  & 0 & 0 & 0 & 0 & \beta  \\
 0 & 0 & \alpha +\beta  & 0 & \beta  & 0 & \beta  & 0 & 0 \\
 0 & -\alpha  & 0 & 0 & 0 & 0 & 0 & 0 & -\alpha  \\
 0 & 0 & 0 & 0 & \alpha  & 0 & 0 & 0 & 0 \\
 0 & 0 & 0 & 0 & 0 & \alpha  & 0 & 0 & 0 \\
 0 & 0 & -\alpha  & 0 & -\alpha  & 0 & 0 & 0 & 0 \\
 0 & 0 & 0 & 0 & 0 & 0 & 0 & \alpha  & 0 \\
 0 & 0 & 0 & 0 & 0 & 0 & 0 & 0 & \alpha  \\
\end{pmatrix}.
\end{equation} This is obtained by choosing a three dimensional representation of the partition algebra generators 
\begin{equation}\label{eq:partitiongene3}
    f_i = \frac{\mathbb{1}+Z_i +Z_i^2}{3}, ~~~ f_{i+\frac{1}{2}}= \mathbb{1}+X_i^2 X_{i+1}+ X_i X_{i+1}^2,
\end{equation}
where $X$ and $Z$ are the $\mathbb{Z}_3$ {\it shift} and {\it clock} matrices 
\begin{equation}
X = \begin{pmatrix}
 0 & 0 & 1 \\
 1 & 0 & 0 \\
 0 & 1 & 0 \\
\end{pmatrix}; ~~ 
Z = \begin{pmatrix}
 1 & 0 & 0 \\
 0 & \omega  & 0 \\
 0 & 0 & \omega ^2 \\
\end{pmatrix}; ~~ \text{where} ~~\omega= e^{\frac{2 \pi \mathrm{i}}{3}}.
\end{equation}
Also note that $\big(f_{i+\frac{1}{2}}\big)^2 = 3 ~ f_{i+\frac{1}{2}}$.

It is possible to compare this operator and others built in this manner with those examined in \cite{hietarinta1993upper,hietarinta2024solutions}. These works examine 9 by 9 solutions that either meet the `additive charge conservation' (ACC) requirement or are in the upper triangular form\footnote{Under the ACC condition $R_{ij,kl}=0$ if $i+j\neq k+l$.}. It is important to note that the answer in \eqref{eq:partition-9by9} does not satisfy either of these conditions and hence is likely to be inequivalent to previous 9 by 9 solutions. However, further investigation is necessary to fully comprehend the algebraic solutions and, more broadly, the classification.
    \item It is possible to construct both local and non-local integrable models using the spectral parameter dependent 4 by 4 solutions, depending on whether the $R$-matrices are regular. An algebraic method similar to the one employed in this study can be applied to this problem \cite{de2019classifying}. Furthermore, this method will assist us in building and analyzing higher spin chains in a systematic way.
    \item Hietarinta's work also has non-invertible constant 4 by 4 YBOs. Twelve of these classes are described by him in \cite{HIETARINTA-PLA}. An algebraic construction of these might also be intriguing. For instance, we may use the partition algebra generators to create a few basic non-invertible YBOs. Consider the operators :
    \begin{eqnarray}
        & f_if_{i+\frac{1}{2}},~ f_{i+\frac{1}{2}}f_i,~ f_{i+1}f_{i+\frac{1}{2}},~ f_{i+\frac{1}{2}}f_{i+1} & \nonumber \\
        & f_i,~ f_{i+1},~ f_if_{i+1},~ f_{i+\frac{1}{2}}   & \nonumber \\
        & P_if_i,~ f_iP_i. & 
    \end{eqnarray} All of the aforementioned are algebraic solutions or representation independent solutions to the braided YBE \eqref{eq:YBEbraided}. Algebraic YBOs that are not invertible and do not have the factorized form include
    \begin{eqnarray}
       & f_if_{i+1}f_{i+\frac{1}{2}} - f_{i+\frac{1}{2}}f_if_{i+1} & \nonumber \\
      & \alpha~f_if_{i+\frac{1}{2}}f_{i+1} + \beta~f_{i+1}f_{i+\frac{1}{2}}f_i, & 
    \end{eqnarray} among many others. In the two dimensional representation these solutions are not equivalent to Hietarinta's non-invertible 4 by 4 YBO's \cite{hietarinta1993-JMP-Long}. We can also obtain some of Hietarinta's non-invertible YBO's algebraically. For instance consider a rank 3 and a rank 2 solution with no parameters
    \begin{eqnarray}
        H0,5 = \begin{pmatrix}
            1 & 0 & 0 & 0 \\
            0 & 1 & 0 & 0 \\
            0 & 0 & 1 & 0 \\
            0 & 0 & 0 & 0
        \end{pmatrix},~~H0,6 = \begin{pmatrix}
            1 & 1 & 1 & 0 \\
            0 & 0 & 0 & 0 \\
            0 & 0 & 0 & 0 \\
            0 & 0 & 0 & 1
        \end{pmatrix}.
    \end{eqnarray}
    The $H0,5$ is obtained from the algebraic solution
    \begin{eqnarray}
        \tilde{R}_{ij} = \mathbb{1} - A_iA_j
    \end{eqnarray}
    when $A^2=A$. We take a two dimensional representation for $A = \begin{pmatrix}
        0 & 0 \\
        0 & 1
    \end{pmatrix}$ to obtain the $H0,5$ class.
    The $H0,6$ solutions is contained in the algebraic operator
     \begin{eqnarray}
        \tilde{R}_{ij} = \mathbb{1} - A_iB_j - B_iA_j,
    \end{eqnarray}
    with $A^2=A$, $B^2=B$ and $AB=0$. The two dimensional representation given by 
    \begin{equation}
A = \begin{pmatrix}
    0 & a\\
    0 & 1
\end{pmatrix}~, ~~ B= \begin{pmatrix}
    1 & 0\\
    0 & 0
\end{pmatrix},
\end{equation}
reproduces the $H0,6$ class.
    The other non-invertible alternatives require a more thorough investigation, which is saved for a later project. Despite not being invertible, these YBOs can be utilized to build integrable models via Baxterization \cite{Jones1990}. 
\end{enumerate}

%%%%%%%%%%%%%%%%%%%%%%%%%%%%%
\section*{Acknowledgments}
%\label{sec:}
%%%%%%%%%%%%%%%%%%%%%%%%%%%%%
We thank Jarmo Hietarinta for valuable comments on the manuscript. VKS and PP thank the organizers of the conference ``Quantum Information and Matter'', May 27-May 31, 2024 at NYU Abu Dhabi, where part of this work was developed. SM and PP thank Indrajit Jana for useful discussions.

VK is funded by the U.S. Department of Energy, Office of Science, National Quantum Information Science Research Centers, Co-Design Center for Quantum Advantage ($C^2QA$) under Contract No. DE-SC0012704. $C^2QA$ led the research through discussions and idea generations, and participated in preparing and editing the draft. The work of VKS is supported by ``Tamkeen under the NYU Abu Dhabi Research Institute grant CG008 and  ASPIRE Abu Dhabi under Project AARE20-336''.

%%%%%%%%%%%%%%%%%%%%%%%%%
%\appendix
%%%%%%%%%%%%%%%%%%%%%%%%%

%%%%%%%%%%%%%%%%%%%%%%%%%%%%%%%%%%%%%%%%%%%%%%%%%%
%\section{Diagram algebras}
%\label{app:diagramAlgebras}
%%%%%%%%%%%%%%%%%%%%%%%%%%%%%%%%%%%%%%%%%%%%%%%%%

\bibliographystyle{acm}
\normalem
\bibliography{refs}

\end{document}